\documentclass[12pt, fleqn]{article}
\usepackage[cp1251]{inputenc}
\usepackage{latexsym,amsfonts,amssymb}
\usepackage{graphicx}

\usepackage{amsbsy}
\usepackage{amsmath}
\usepackage{epsf}
\usepackage{cite}

\newtheorem{theorem}{Theorem}

\newtheorem{remark}{Remark}
\newcommand{\bt}{\begin{theo}}
\newcommand{\et}{\end{theo}}
\newcommand{\bd}{\begin{displaymath}}
\newcommand{\ed}{\end{displaymath}}

\newcommand{\be} {\begin{equation}}
\newcommand{\ee} {\end{equation}}
\newcommand{\ba} {\begin{array}}
\newcommand{\ea} {\end{array}}
\newcommand{\bea}{\begin{eqnarray}}
\newcommand{\eea} {\end{eqnarray}}

\begin{document}

\begin{center}
 {\Large \bf Exact solutions of
 a  mathematical model 
 \\
 \vspace{0.3cm}
   describing  competition and co-existence   \\ \vspace{0.3cm} of different language speakers}
\medskip

{\bf Roman Cherniha \footnote{\small  Corresponding author. E-mail: r.m.cherniha@gmail.com}}
  {\bf and  Vasyl' Davydovych \footnote{\small  E-mail:davydovych@imath.kiev.ua }}
 \\
{\it ~Institute of Mathematics,  National Academy
of Sciences  of Ukraine,\\
 3, Tereshchenkivs'ka Street, Kyiv 01004, Ukraine
}\\
 \end{center}

 \begin{abstract}
The known three-component reaction-diffusion system modeling  competition and co-existence  of different language speakers is under study.
A modification of this system is proposed, which  is examined by Lie
symmetry method; furthermore exact solutions in the form of  traveling
fronts are constructed and   their properties are identified.
 Plots of the traveling fronts  are presented and the
 relevant interpretation describing the language shift occurred in
  Ukraine during the Soviet times is suggested.
\end{abstract}

\emph{Keywords:}
Reaction-diffusion system, Lie symmetry, exact solution, traveling front, \\ community  of language speakers.

\section{\bf  Introduction} \label{sec:1}

 It is well known at least 100 years that many processes arising in physics, chemistry, ecology etc. can be adequately described only by nonlinear partial differential (integro-differential, functional-differential) equations (see, e.g., an extensive discussion on this matter in Chapter 1 of \cite{ch-se-pl-2018}).
 During the second half of the last century, one may note also  a rapidly growing number of papers devoted to applications of  nonlinear partial differential  equations for mathematical modeling in life sciences (see, e.g., the classical book \cite{mur2003}, the recent monographs \cite{ku-na-ei-16, ch-da-2017} and  references therein).

 On the other hand, the rigorous mathematical models came to social sciences and humanities only recently. In particular,  papers devoted to rigorous mathematical modeling interaction of  communities  (populations) of different language speakers were published only during the last two decades \cite{abram-str-03,patr-lepp-04,pin-rom-06,kandler-08,kandler-stee-08,kandler-10, kandler-17}. These models are based on nonlinear differential equations  of reaction-diffusion type.

 Here we study the nonlinear mathematical model describing interaction of three communities of language speakers proposed in \cite{kandler-10}.
  The model is governed by three nonlinear reaction-diffusion (RD) equations, which have the following form  in the one-dimensional approximation (there are some misprints in \cite{kandler-10}, which  are corrected here)

\begin{equation}\label{1-1}\begin{array}{l}  u_t = \lambda_1 u_{xx}+a_1u\left(1-\frac{u}{K-(v+w)}\right)-c_{31}uw+c_{12}uv,\\
 v_t =
\lambda_2  v_{xx}+ a_2v\left(1-\frac{v}{K-(u+w)}\right)+\left(c_{13}+c_{31}\right)uw-\left(c_{12}u+c_{32}w\right)v,\\
w_t = \lambda_3
w_{xx}+a_3w\left(1-\frac{w}{K-(u+v)}\right)-c_{13}uw+c_{32}vw.
\end{array}\end{equation}

This model (of course, one needs to supply the relevant initial and boundary conditions) describes interaction of three communities of language speakers.
 Functions $u(t,x)$ and  $w(t,x)$  describe frequencies
   of monolingual speakers, i.e. they speak always (or almost always) native language. Function  $v(t,x)$  stands for community of  speakers, who fluently speak both languages and use each language depending on circumstances.
 Time derivatives $u_t,  v_t$ and $w_t$ indicate the rate of change in these frequencies, while the space-derivatives describe mobility (diffusion) in space of speakers.
The second terms in each equation of (\ref{1-1}) are some generalization of a standard logistic terms arising in many well known biological  models including the famous Fisher equation \cite{fi-37} and the diffusive Lotka--Volterra system (DLVS) for interacting species (see, e.g., \cite{mur2003,ch-da-2017}).
The constant $K$ (like in  the logistic terms) means the carrying capacity of environment and defines an upper size of all three communities of speakers, i.e. it is assumed that
  $u+v+w < K.$

The language shift (a process whereby speakers of a community abandon their native language in favor of another)  of some numbers of monolingual speakers to bilingual those
is described by the terms  $c_{31}uw$ and  $c_{13}uw$.
It can be noted that the language shift leads to growing the bilingual community
(provided any other forces are absent).

On the other hand,  the terms   $c_{12}uv$ and  $c_{32}vw$  describe  an opposite tendency, when bilingual  tends to be  monolingual. It occurs, for example, in the case of the state politics leading to the lower status of one language comparing with another. The real example is the  Russification in Ukraine during the Soviet period when a few millions of Ukrainians completely  switched to the Russian  language (actually, the main aim of paper \cite{kandler-10} is to study mathematically Anglicization in Scotland). The coefficients $c_{12}$ and  $c_{32}$ represent the likelihood of bilingual speakers then becoming monolingual  in community $u$  and $w$, respectively. Notable, the inequality $c_{12} < c_{32}$ (in particular, if  $c_{12} << c_{32}$ then one puts $c_{12}=0$) takes place if the language of community $u$ is under pressure.

In paper  \cite{kandler-10},   the RD system  (\ref{1-1}) was used
 in order to model the Anglicization process in Scotland during the 20th century. As a result, percentages of  Gaelic  speakers in all parts of Scotland  decreased drastically. However, there is no any mathematical analysis of the governing equations therein, while those were solved numerically (with the relevant boundary and initial conditions)  in order to show a good correspondence between the numerical solutions and data from successive censuses.

In this paper, a modification of the RD system  (\ref{1-1}) is studied by analytical methods
and  a plausible interpretation of the mathematical results obtained is provided.
The main results are presented in Section~\ref{sec:2}. Firstly,  a  modification
of the system in question  is proposed, secondly, Lie symmetries and a variety of exact solutions (traveling waves) are found.  In Section~\ref{sec:3}, properties of the exact solutions obtained are under examination, in particular,  the coefficient restrictions leading to the exact solution, which describes qualitatively   the language shift occurred in  Ukraine during the Soviet times, are derived. Finally, some conclusions are   presented  and a future work  is announced   in the last section.

\section{Main results}\label{sec:2}

The RD system  (\ref{1-1}) contains fractional nonlinearities and it is a very difficult task   to solve analytically  such type  systems.
Having this in mind, we propose here a simpler system under biologically motivated restrictions. Our idea is to reduce fractional nonlinearities to quadratic those.  It can be noted that  the  fractional nonlinearities arising  in  (\ref{1-1}) are a  direct generalization of those introduced in the earlier work  \cite{kandler-stee-08}. In that work,  it is assumed that speakers of both languages have  the common carrying capacity $K$. We  think that this assumption is not well-founded because   language  of a specified speaker  is usually  related to his/her nationality. So, one cannot claim that different nationalities have the  same  carrying capacities. Moreover, the  so-called  standard model for two competing languages  \cite{pin-rom-06} does not use  such assumption. The basic model in  \cite{pin-rom-06} contains the standard logistic terms arising in many biologically motivated models  (see, e.g.  \cite{mur2003, ch-da-2017}).
 Taking into account the above justification,   we  can  replace the fractional nonlinearities by logistic terms, which also restrict  unbounded growth of these communities. It means that the terms
$\frac{u}{K-(v+w)}, \  \frac{v}{K-(u+w)}, $  and $  \frac{w}{K-(v+u)}$ are replaced  by $\frac{u}{K_1}, \  \frac{v}{K_2}$,  and  $\frac{w}{K_3}$, respectively.   As a result,  we obtain the following modification of system~(\ref{1-1})
 \begin{equation}\label{1-2}\begin{array}{l}  u_t = \lambda_1u_{xx}+a_1u\left(1-\frac{u}{K_1}\right)-c_{31}uw+c_{12}uv,\\
 v_t =
\lambda_2 v_{xx}+ a_2v\left(1-\frac{v}{K_2}\right)+\left(c_{13}+c_{31}\right)uw-\left(c_{12}u+c_{32}w\right)v,\\
w_t = \lambda_3
w_{xx}+a_3w\left(1-\frac{w}{K_3}\right)-c_{13}uw+c_{32}vw,
\end{array}\end{equation}
which contains only quadratic nonlinearities.
Hereafter we assume that the  coefficients  $\lambda_i, \ a_i$ and $K_i$
($i=1,2,3$) are positive, while all other are nonnegative (i.e., some of them can be zero).

The nonlinear RD system (\ref{1-2}) can be simplified using the
following re-scaling  of the variables
\begin{equation}\nonumber u\rightarrow K_1u, \
v\rightarrow K_2v, \ w\rightarrow K_3w, \
t\rightarrow\frac{1}{a_2}\,t, \
x\rightarrow\sqrt{\frac{\lambda_2}{a_2}}\,x\end{equation}
and introducing new notations
\[\begin{array}{l}\alpha_1=\frac{c_{31}K_3}{a_2}, \ \alpha_2=\frac{c_{12}K_2}{a_2},
 \ \alpha_3=\frac{c_{13}K_1}{a_2}, \ \alpha_4=\frac{c_{32}K_2}{a_2},
 \\
\beta_1=\frac{a_1}{a_2}, \ \beta_3=\frac{a_3}{a_2},\
\kappa_1=\frac{K_3}{K_2}, \ \kappa_2=\frac{K_1}{K_2}, \
d_1=\frac{\lambda_1}{\lambda_2}, \
d_3=\frac{\lambda_3}{\lambda_2}.\end{array}\] Thus, system (\ref{1-2})
is reduced to the equivalent form
 \begin{equation}\label{1-4}\begin{array}{l}
 u_t = d_1u_{xx}+\beta_1u\left(1-u\right)-\alpha_1uw+\alpha_2uv,\\
 v_t =
v_{xx}+ v\left(1-v\right)+\left(\kappa_1\alpha_3+\kappa_2\alpha_1\right)uw-\left(\kappa_2\alpha_2u+\kappa_1\alpha_4w\right)v,\\
w_t = d_3 w_{xx}+\beta_3w\left(1-w\right)-\alpha_3uw+\alpha_4vw.
\end{array}\end{equation}

Notably, system (\ref{1-4}) with $\alpha_1=\alpha_3=0$
is a particular case of  the well-known DLVS,
which describes a large  number of processes in biology and chemistry (see, e.g.,
\cite{mur2003,ch-da-2017} and references cited therein). However the above restriction  is equivalent to  $c_{13}=c_{31}=0$ in (\ref{1-2}),
what contradicts to the basic restrictions in the model (see interpretation of the terms   $c_{31}uw$ and $c_{13}uw$). Thus, hereafter we assume that
$c_{13}^2+c_{31}^2\neq0 \Leftrightarrow \alpha_{1}^2+ \alpha_{3}^2\neq0 $, i.e., system (\ref{1-4}) is not equivalent to DLVS.

 It is well known that there is no general theory of integrating nonlinear partial differential equations at the present time and it is very unlikely that one will be developed soon.
The most effective methods for constructing {\it particular exact solutions}
of nonlinear differential equations of reaction-diffusion
type are the classical Lie method and its various generalizations (see, e.g., the recent
monographs \cite{bl-anco-10, arrigo15, ch-se-pl-2018} for more details). Here we apply the classical Lie method and the  so-called tanh method \cite{malfliet-her-96,malfliet-04,waz-08}.

\begin{theorem} The nonlinear system (\ref{1-4}) for any set of specified nonnegative coefficients
with the additional restrictions
$d_1d_3\kappa_1\kappa_2\not=0$ and $\alpha_1^2+\alpha_3^2\not=0$
is invariant only  with respect to the time and space translations generated by Lie symmetries
 \begin{equation}\label{1-5}  P_t = \frac{\partial}{\partial t}, \ P_x = \frac{\partial}{\partial x}. \end{equation}
\end{theorem}

The proof is based on  application of the well known Lie's algorithm to system (\ref{1-4}) and is reduced to examination of several cases depending on values of  the coefficients   arising in the  system. We  omit  here the relevant calculations. Notably, a detailed proof is presented in our recent paper
 \cite{ch-dav-2019} for a similar (but inequivalent!) three-component  system.

 \begin{remark} In contrast to the three-component DLVS, which admits some nontrivial Lie symmetries (provided its coefficients are correctly-specified)   \cite{che-dav2013, ch-da-2017}, the RD system
   (\ref{1-4}) possesses a poor symmetry.
\end{remark}

It is well known that the Lie symmetries
 (\ref{1-5}) generate only two inequivalent substitutions (following the classical Sophus Lie papers, the terminology 'ansatz' is often used), which reduce  system (\ref{1-4})  to the relevant systems of ordinary differential equations (ODEs).
 The first ansatz does not depend on the space variable $x$, hence one leads only to  time-dependent    solutions. Here we are not interested in such type solutions because their realistic interpretation is questionable.

The second ansatz follows from the linear combination $P_t +\mu P_x$ of the Lie symmetries (\ref{1-5}) and has the form
  \begin{equation}\label{1-6} u=U(\omega), \  v=V(\omega),\  w=W(\omega), \ \omega=x-\mu t, \  \mu \in
 \mathbf{R}.
 \end{equation} Here
  $U, \ V$ and $W$ are new unknown functions.
  Solutions of form (\ref{1-6}) is often called plane wave  solutions (traveling waves).
  From the applicability point of view, the most interesting those are {\it traveling fronts}, i.e. solutions (\ref{1-6}), which are
bounded and nonnegative. A huge number of papers is devoted to construction of  traveling fronts for nonlinear PDEs, especially for scalar reaction-diffusion (with/without  convection term). A majority of traveling fronts for such type equations are presented in the monograph \cite{gi-ke-04} (see also the handbook \cite{polya-zait-2012}).

 In the case of nonlinear RD systems, the progress is rather  modest. To the best of our knowledge, an essential progress is derived only in the case of DLVS. Several traveling fronts are constructed in \cite{rod-mimura-2000,ch-du-04,hung2012,ch-da-2017} for the two-component DLVS and in \cite{hung,hung2011} for the three-component DLVS.

So, our aim is to find traveling fronts for system  (\ref{1-4}).
  Substituting ansatz (\ref{1-6}) into system (\ref{1-4}), one obtains
\begin{equation}\label{2-1}\begin{array}{l}d_1U''+\mu\, U'+\beta_1U\left(1-U\right)-\alpha_1UW+\alpha_2UV=0,\\
V''+\mu\, V'+V\left(1-V\right)+\left(\kappa_1\alpha_3+\kappa_2\alpha_1\right)UW-\left(\kappa_2\alpha_2U+\kappa_1\alpha_4W\right)V=0,\\
d_3W''+\mu\, W'+\beta_3W\left(1-W\right)-\alpha_3UW+\alpha_4VW=0.
 \end{array}\end{equation}
System (\ref{2-1}) is  three-component system of  nonlinear
second-order ODEs. Although this system is simpler
object than the original RD system (\ref{1-4}), we can say nothing about its integrability because even the similar system obtained by reducing of the two-component DLVS has been not solved in  \cite{rod-mimura-2000,ch-du-04,hung2012,ch-da-2017}.
In order to find particular solutions of (\ref{2-1}), we start from the steady-state points. Obviously that steady-state points of (\ref{2-1}) coincide with the stationary (homogenous)   those of the RD system (\ref{1-4}) and can be easily calculated by solving algebraic equations. Assuming $u_0v_0w_0=0$, the full list of  steady-state points are as follows
\begin{equation}\label{2-1*}\begin{array}{l}(0,0,0), \ (0,1,0), \ (0,0,1), \ (1,0,0), \\
 \left(\frac{\beta_1+\alpha_2}{\beta_1+\kappa_2\alpha_2^2},
\frac{\beta_1(1-\kappa_2\alpha_2)}{\beta_1+\kappa_2\alpha_2^2},0\right),
\ \left(0,
\frac{\beta_3(1-\kappa_1\alpha_4)}{\beta_3+\kappa_1\alpha_4^2},\frac{\beta_3+
\alpha_4}{\beta_3+\kappa_1\alpha_4^2}\right).\end{array}\end{equation}
  Obviously there are also  steady-state points  $(u_0,v_0,w_0)$, where
  $u_0v_0w_0\not=0$, however we prefer examine this case elsewhere.
  Notably, the 3rd and 4th points, similarly to 5th and 6th those, are equivalent because the first and third equations of system  (\ref{2-1}) have the same structure. So, without loss of generality we may say that there are only four essentially different points in (\ref{2-1*}).

Typically, each traveling front possesses the following property: such solution connects two steady-state points provided $\omega \rightarrow \pm
\infty$. We were able to identify the relevant traveling fronts  in the cases listed below.

\textbf{Case 1}.
$(U_0,V_0,0)=\left(\frac{\beta_1+\alpha_2}{\beta_1+\kappa_2\alpha_2^2},
\frac{\beta_1(1-\kappa_2\alpha_2)}{\beta_1+\kappa_2\alpha_2^2},0\right)$
(as $\omega \rightarrow -\infty$) and $(0,0,1)$ (as $\omega
\rightarrow +\infty$).

\textbf{Case 2}. $(U_0,V_0,0)$ (as $\omega \rightarrow -\infty$)
and $(0,0,0)$ (as $\omega \rightarrow +\infty$).

\textbf{Case 3}.  $(1,1,0)$ (as $\omega \rightarrow -\infty$) and $(0,1,0)$ (as $\omega \rightarrow +\infty$).
This case occurs provided the additional restriction $\alpha_2=0$ takes place.

Let us consider \textbf{Case 1} and use the  tanh method.
To the best of our knowledge paper \cite{malfliet-her-96} is one of the earliest works devoted to the tanh method (there are a lot recent papers, see, e.g. \cite{waz-08,abdelkawy} and papers cited therein). However, it can be noted that there are not many papers devoted to application of this method to nonlinear systems of PDEs. The method is essentially based at the ad hoc
ansatz   \cite{malfliet-her-96}
\begin{equation}\label{7*} u(t,x)=U(\omega)=\sum\limits_{i=0}^{N}\gamma_iY^i,
\end{equation} where $Y=\tanh \omega$. The highest power $N$ should be determined by balancing the highest degree terms in $Y$, upon substitution of  ansatz (\ref{7*}) into the equation in question. When one makes balancing, the known relation $\left(\tanh \omega\right)'=1-\tanh^2 \omega$ is essentially used. Typically direct calculation show that $N\leq2$ for the second-order PDEs. So, we obtain ansatz
\begin{equation}\nonumber u(t,x)=\gamma_0+\gamma_1\tanh \omega+\gamma_2\tanh^2 \omega.
\end{equation}

 Having the correctly-specified $N$, unknown parameters $\gamma_i$  can be easily calculated (some of them are arbitrary constants). Of course, it often happens that $N=0$, therefore a trivial solution is only obtained. So, the tanh method is not applicable to a wide range of nonlinear equations. It turns out that this technique works in the case of system (\ref{2-1}).

Thus, using   ansatz (\ref{7*}),  we  may  look  for traveling fronts of the form
\begin{equation}\label{2-2}\begin{array}{l}
 U= \sigma_1\left(1-\tanh \omega \right)^{n_1}, \
V=\sigma_2\left(1-\tanh  \omega \right)^{n_2}, \
W=1-\sigma_3\left(1-\tanh  \omega \right)^{n_3},
\end{array}\end{equation} where  $\sigma_i$  and  $n_i$ ($i=1,2,3$) are real and natural numbers, respectively. Since the exact solution of the form (\ref{2-2}) connects
steady-state points $(U_0,V_0,0)$ and $(0,0,1)$,  one immediately obtains the
sigma-s values
\begin{equation}\label{2-3}\sigma_1=\frac{\beta_1+\alpha_2}{2^{n_1}\left(\beta_1+\kappa_2\alpha_2^2\right)},
\
\sigma_2=\frac{\beta_1(1-\kappa_2\alpha_2)}{2^{n_2}\left(\beta_1+\kappa_2\alpha_2^2\right)},\
 \sigma_3=\frac{1}{2^{n_3}}.\end{equation}

Substituting (\ref{2-2}) into system (\ref{2-1}) and
taking into account (\ref{2-3}), one can determine sufficient conditions for the coefficients $n_i$ when the traveling fronts can be found explicitly.

Omitting the relevant calculations, we present only the result. So, system (\ref{1-4}) has  the exact solution
\begin{equation}\label{2-4}\begin{array}{l} u=\frac{6d_1}{\beta_1}\big(1 -\tanh(x-\mu t)\big)^2, \medskip\\
v=\frac{24d_1-\beta_1}{2 \alpha_2}\big(1 -\tanh(x-\mu t)\big),  \medskip\\
w=\frac{1}{2} +\frac{1}{2}\tanh(x-\mu t)
\end{array}\end{equation}
 provided its coefficients satisfy
the restrictions:
\begin{equation}\label{2-5}\begin{array}{l}\alpha_1=16d_1-4 \mu
+\beta_1,\ \alpha_3=\frac{d_3 \beta_1}{3d_1},\ \kappa_1=\frac{5-2
\mu }{\alpha_4}, \
\kappa_2=\frac{\beta_1\left(\alpha_2+\beta_1-24d_1\right)}{24d_1
\alpha_2^2},\\ \beta_1=\frac{2 \alpha_2^2
\left(\alpha_4+(2\mu-5)d_3\right)}{(10 d_1-\mu
+2\alpha_2)\alpha_4}+24d_1-\alpha_2,\ \beta_3=\frac{2
(2d_3-\mu)\alpha_2+ (\beta_1-24d_1)\alpha_4}{\alpha_2}.
\end{array}\end{equation}
 The second   exact solution
\begin{equation}\label{2-6}\begin{array}{l}u=\frac{\beta_1+\alpha_2}{4\left(\beta_1+\kappa_2\alpha_2^2\right)}
 \big(1 -\tanh(x-10 t)\big)^2, \medskip\\
v=\frac{\beta_1(1-\kappa_2\alpha_2)}{4\left(\beta_1+\kappa_2\alpha_2^2\right)}\big(1 -\tanh(x-10 t)\big)^2,\medskip\\
w=1-\frac{1}{4}\big(1 -\tanh(x-10 t)\big)^2,
\end{array}\end{equation} was constructed  provided the coefficients of system  (\ref{1-4}) satisfy
the restrictions:
\begin{equation}\label{2-7}\begin{array}{l}d_1=1, \ d_3=1, \ \alpha_1=\beta_1-24,\medskip \\
\kappa_1=\frac{24\alpha_2\kappa_2+23\beta_1-\left(\beta_1-24+24
\alpha_2\right)\beta_1\kappa_2}{(\alpha_3-\alpha_4)\beta_1+\left(\alpha_3+\alpha_4
\beta_1 \kappa_2\right)\alpha_2},\
\beta_3=\frac{(\alpha_3-\alpha_4-24)\beta_1-24\kappa_2\alpha_2^2+\left(\alpha_3+\alpha_4
\beta_1 \kappa_2\right)\alpha_2}{\beta_1+\kappa_2\alpha_2^2}.
\end{array}\end{equation}
It is easily   seen that  the traveling front
(\ref{2-4})
 is more general than (\ref{2-6}), since   its velocity $\mu$ is not fixed.

In   \textbf{Case 2}, taking into account the corresponding steady-state points,  we are looking for the  traveling fronts in the form
\begin{equation}\label{2-8}\begin{array}{l}
 U= \frac{\beta_1+\alpha_2}{2^{n_1}\left(\beta_1+\kappa_2\alpha_2^2\right)}\left(1-\tanh \omega \right)^{n_1}, \medskip\\
V=\frac{\beta_1(1-\kappa_2\alpha_2)}{2^{n_2}\left(\beta_1+\kappa_2\alpha_2^2\right)}\left(1-\tanh  \omega \right)^{n_2}, \medskip\\
W=\sigma\left(1-\tanh^2  \omega \right),
\end{array}\end{equation} where $\sigma$ is an unknown positive constant. Substituting (\ref{2-8})
 into system (\ref{2-1}) and  making the corresponding calculations, we arrive at the exact solution
\begin{equation}\label{2-9}\begin{array}{l}
 u=\frac{17-16d_1+\alpha_2}{68-64d_1+4\kappa_2\alpha_2^2}\left(1-\tanh \left(x-\frac{17}{4}\,t\right) \right)^2, \medskip\\
v=\frac{(17-16d_1)(1-\kappa_2\alpha_2)}{68-64d_1+4\kappa_2\alpha_2^2}\left(1-\tanh \left(x-\frac{17}{4}\,t\right) \right)^2, \medskip\\
w=\frac{17-40d_1}{4\alpha_1}\left(1-\tanh^2
\left(x-\frac{17}{4}\,t\right) \right).
\end{array}\end{equation}
 The  traveling front (\ref{2-9}) satisfies system (\ref{1-4}) if  the
coefficient restrictions
\begin{equation}\label{2-10}\begin{array}{l}\beta_1=17-16d_1,\ \beta_3=\frac{17-8d_3}{2},\
\alpha_4=\frac{16d_1(17-\alpha_3)+(17+
 \alpha_2)\alpha_3-17\left(17+\kappa_2\alpha_2^2\right)}{(17-16d_1)(1-\kappa_2\alpha_2)},\medskip\\
d_3=\frac{17}{8}-\frac{17\alpha_1}{12\alpha_1+80d_1-34},\
\kappa_1=\frac{\alpha_1}{17(17-40d_1)}\frac{391-368d_1-(289 - 952
d_1 + 640 d_1^2 + 408 \alpha_2 -
     408 d_1 \alpha_2) \kappa_2}{17-16d_1+\kappa_2\alpha_2^2}
\end{array}\end{equation} are satisfied.

  Finally, in \textbf{Case 3}, the exact solutions of system (\ref{2-1}) were prescribed to have the form
\begin{equation}\nonumber\begin{array}{l}
 U= \frac{1}{2^{n_1}}\left(1-\tanh \omega \right)^{n_1}, \
V=1+\sigma_2\left(1-\tanh^2  \omega \right), \
W=\sigma_3\left(1-\tanh^2  \omega \right).
\end{array}\end{equation}
After the relevant calculations, the traveling front
\begin{equation}\label{2-12}\begin{array}{l}
 u= \frac{1}{4}\left(1-\tanh \left(x-\frac{\alpha_3}{4}\,t\right) \right)^2, \medskip\\
v=1+\frac{24 - \alpha_3}{2 (\alpha_3-8)}\left(1-\tanh^2 \left(x-\frac{\alpha_3}{4}\,t\right) \right), \medskip\\
w=\frac{\alpha_3-40d_1}{4\alpha_1}\left(1-\tanh^2
\left(x-\frac{\alpha_3}{4}\,t\right) \right),
\end{array}\end{equation} of the nonlinear  system
(\ref{1-4}) was derived provided the
coefficient restrictions
\begin{equation}\label{2-13}\begin{array}{l}
\alpha_2=0, \ \beta_1 = -16 d_1 + \alpha_3,\ \beta_3 =\frac{2
\alpha_1 \alpha_3 [\alpha_3 - 2 (4 + \alpha_4)]}{(\alpha_3 -8)(40
d_1 + 6 \alpha_1 - \alpha_3)},\ d_3=\frac{\alpha_3 - 2 \alpha_4 - 2
\beta_3}{8},
\medskip\\ \kappa_1=\frac{\alpha_1(\alpha_3 -24)(\alpha_3
-6)}{\alpha_4(\alpha_3 -8)(\alpha_3 -40d_1)}, \
\kappa_2=\frac{\alpha_3(6 - \alpha_3 + 2
\alpha_4)}{\alpha_1(\alpha_3 -6)}\kappa_1,
\end{array}\end{equation} take place.

\begin{remark} In  Cases 1--3 there exist such  sets of the positive  parameters (excepting $\alpha_2=0$  in \textbf{Case~3})
\[d_1,\ d_3,\ \alpha_i, \ \beta_1, \ \beta_2, \ \kappa_1, \ \kappa_2,\]
satisfying the restrictions (\ref{2-5}), (\ref{2-7}), (\ref{2-10})
and  (\ref{2-13})  that three
 components of the exact solutions (\ref{2-4}), (\ref{2-6}),
(\ref{2-9}) and (\ref{2-12}), respectively,  are positive. Thus, all the solutions obtained  are indeed traveling fronts.
\end{remark}

\begin{remark}
It can be easily checked that all the solutions derived above satisfy the zero Neumann conditions at $x \rightarrow \pm \infty$.
In the case of a bounded domain $(A, B) $, one obtains at the boundaries  $u_x \approx 0, \  v_x \approx 0$  and $ w_x \approx 0$  provided  $|A| $   and  $|B| $ are sufficiently large. Such conditions are typical requirements  and, for instance, were used in \cite {kandler-10}.
\end{remark}

\section{Interpretation of   traveling fronts} \label{sec:3}

In this section, we study in detail  exact solution  (\ref{2-4}).
  First  of all, we  answer the question: When  positive coefficients
$d_1, \ d_3, \ \alpha_2, \ \alpha_4$ and $\mu$ lead automatically to
positive values of $\alpha_1, \ \alpha_3, \ \beta_1, \ \beta_3, \
\kappa_1$ and $\kappa_2$ in formulae (\ref{2-5})? It turns out that
some  additional restrictions are needed. The structure of such
restrictions essentially depends on the sign of the parameter $\mu$, i.e. on the traveling front direction.
Thus, one needs to examine separately two cases: \textbf{(i)}~$\mu>0$
and \textbf{(ii)} $\mu<0$.

In Case \textbf{(i)}, one immediately obtains $0<\mu<\frac{5}{2}$  (see the formula for $\kappa_1$ in (\ref{2-5})).
For a simplicity, we assume additionally  $\alpha_2=\alpha_4\equiv\alpha$  and
introduce the notations
 \[F\equiv10 d_1-\mu +2\alpha, \ G\equiv2\mu
d_3-5d_3+\alpha.\]
Substituting these notations into (\ref{2-5}), we
arrive at the system of the  inequalities:
\begin{equation}\label{3-1}\begin{array}{l} FG>0, \
\alpha_1=40d_1-4\mu-\alpha\left(1-2\,\frac{G}{F}\right)>0,\medskip \\
\beta_1=24d_1-\alpha\left(1-2\,\frac{G}{F}\right)>0, \
\beta_3=4d_3-2\mu-\alpha\left(1-2\,\frac{G}{F}\right)>0.
\end{array}\end{equation}
Since all the component of (\ref{2-4}) should be nonnegative (we remind the reader that each component means a frequency of the community speakers),
  the inequality $\beta_1<24d_1$ takes place, which follows from  $V\geq0$.
  Thus, the restriction
$\frac{G}{F}<\frac{1}{2}$  is obtained. It can be also noted that $F>0$ and $G>0$ (the
case $F<0$ and $G<0$ leads to a contradiction).

 In order to satisfy
all the inequalities in (\ref{3-1}), we set \[ G=\varepsilon
\Leftrightarrow \alpha=(5-2\mu)d_3+\varepsilon, \]
where
$\varepsilon>0$ is a sufficiently small parameter. Now the 4th inequality in
(\ref{3-1}) is reduced to the form: \begin{equation}\label{3-2}
d_3\geq\frac{2\mu+\varepsilon}{2\mu-1}, \end{equation} hence $\mu>\frac{1}{2}.$
The 2nd and 3rd those are satisfied provided
\begin{equation}\label{3-3}
40d_1>4\mu+5d_3-2\mu d_3+\varepsilon, \quad  24d_1>5d_3-2\mu
d_3+\varepsilon.
\end{equation}

Now one realizes that the following algorithm
guarantees the positivity of all the coefficients in (\ref{2-5}). Firstly, we
fix any $\mu$ from the interval $\left(\frac{1}{2},\frac{5}{2}\right)$
and a small $\varepsilon$, say $\varepsilon<1$. Secondly, we take any
$d_3$ satisfying (\ref{3-2}) and calculate
$\alpha=(5-2\mu)d_3+\varepsilon.$ Finally, we choose a sufficiently
large $d_1>0$ in order to satisfy  inequalities (\ref{3-3}).
\begin{remark} In the case $\alpha_2=\alpha_4\equiv\alpha$ and $d_1=d_3\equiv
d$, the above  algorithm is simplified to the identification of  the restrictions
$d\geq\frac{2\mu+\varepsilon}{2\mu-1}$ and
$\alpha=(5-2\mu)d+\varepsilon,$ where $\varepsilon>0, \mu \in \left(\frac{1}{2},\frac{5}{2}\right)$.
\end{remark}

Case \textbf{(ii)} is essentially  simpler. In fact,  one immediately
obtains $\alpha_1>0$ and $\kappa_1>0$ in (\ref{2-5}). Assuming
additionally that $\alpha_2=24d_1$ and solving the inequalities
$\beta_1>0$ and $\beta_3>0$ (see (\ref{2-5})), we obtain the restrictions
\begin{equation}\nonumber\begin{array}{l} \alpha_2=24d_1, \ d_3<1,
\ \mu<\frac{d_3}{2(d_3-1)},\medskip\\
(5-2\mu)d_3<\alpha_4<\frac{2}{10d_1-\mu}\left(\mu^2+2(24d_1d_3-d_3-29d_1)\mu-4d_1d_3\right),
\end{array}\end{equation} which guarantee  the positivity of all the
coefficients in (\ref{2-5}).

\begin{figure}[ht]
\centering
 \includegraphics[width=6cm]{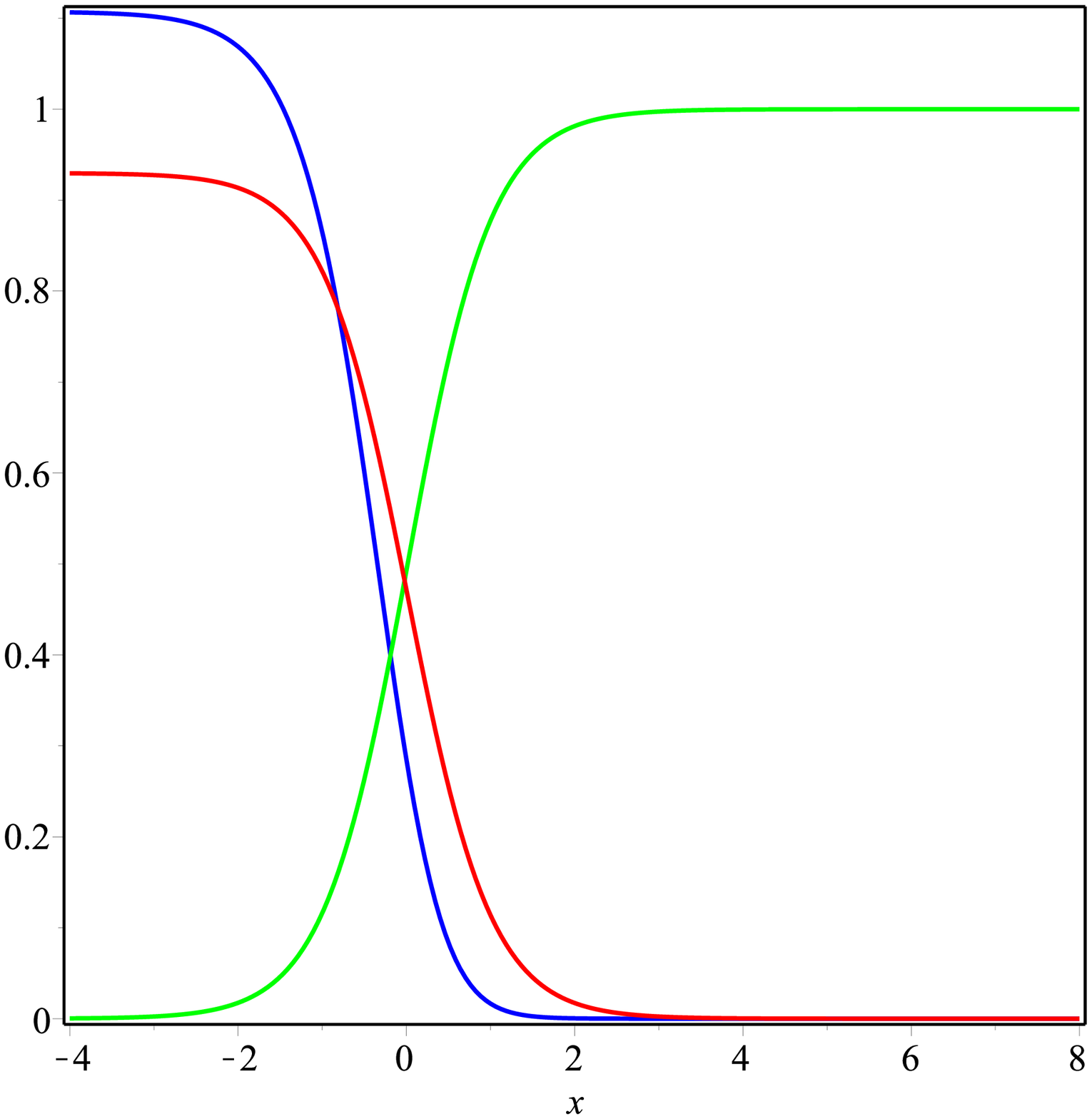}
  \includegraphics[width=6cm]{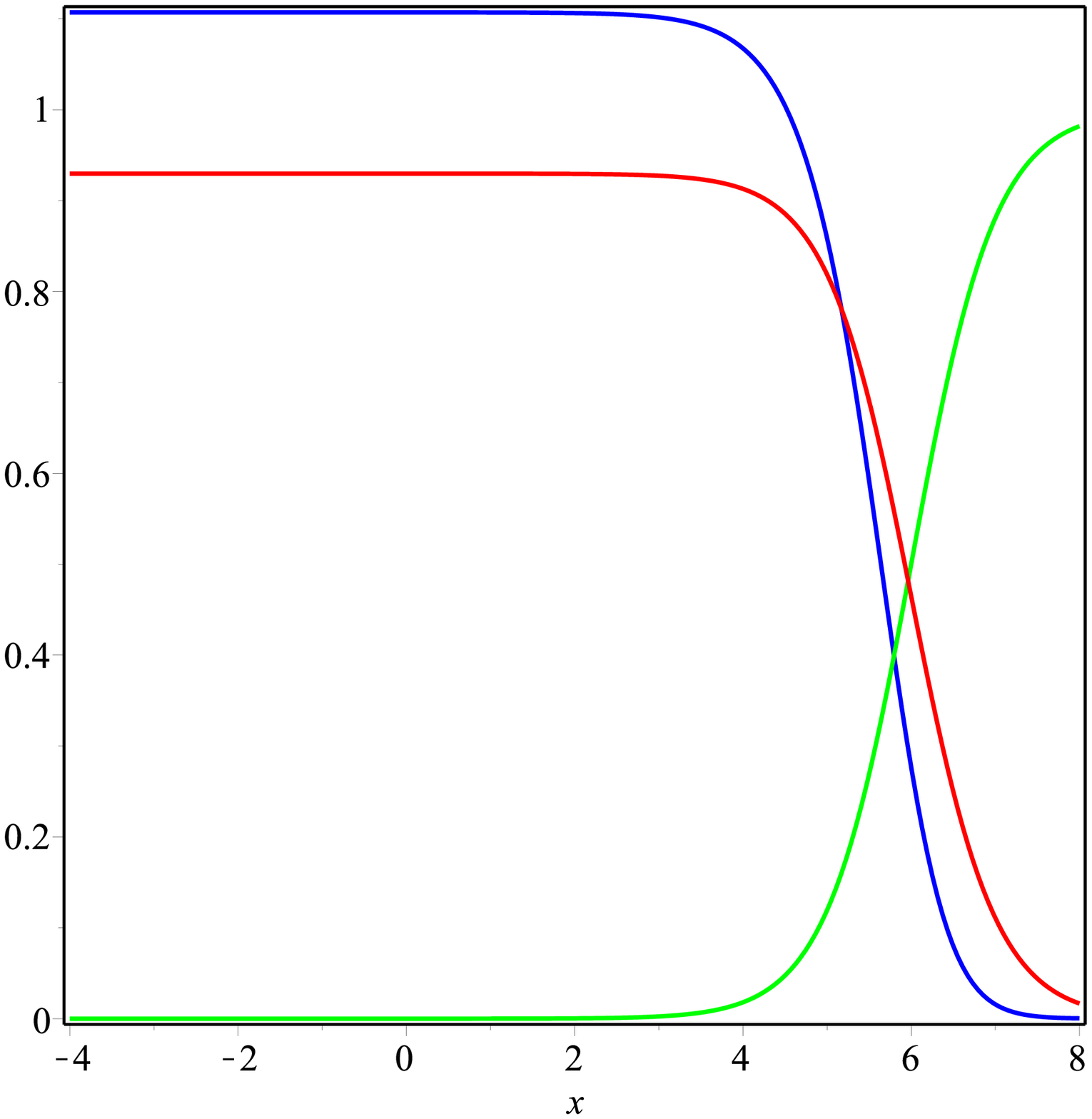}
\caption{Traveling fronts (\ref{2-4}). Curves  represent the functions $u(t_0,x)$ (blue represents the Russian speakers),
$v(t_0,x)$ (red represents the bilingual  speakers) and $w(t_0,x)$ (green  represents the  Ukrainian  speakers)   for the fixed  time $t_0=0.01$ (left) and $t_0=4$ (right) and the parameters
  $\mu=\frac{3}{2}, \ d_1=d_3=2, \ \alpha_2=\alpha_4=5$
   (other parameters are calculated by formulae (\ref{2-5})). }\label{f1}
\end{figure}

\begin{figure}[ht]
\centering
  \includegraphics[width=6cm]{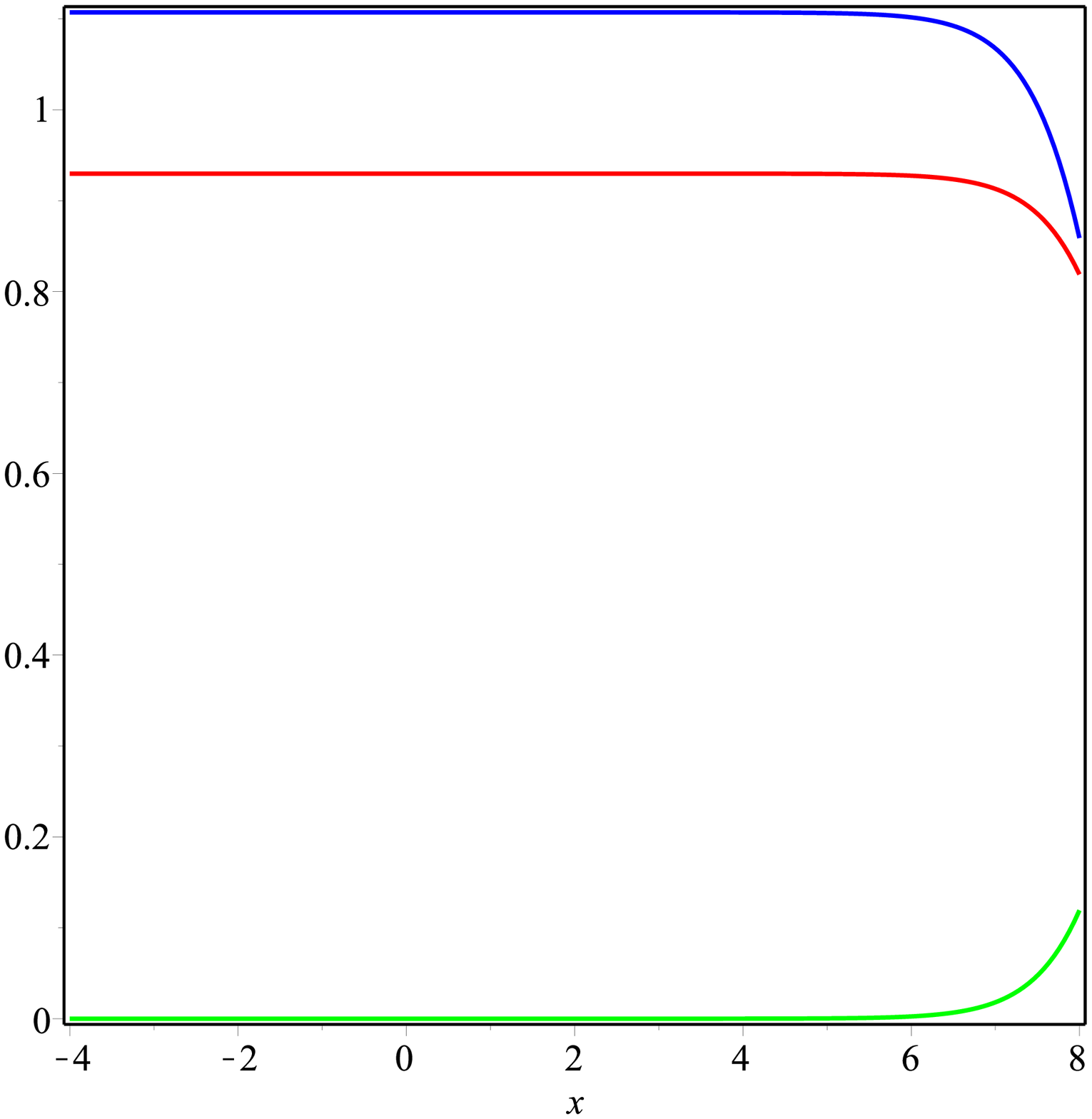}
\caption{Traveling fronts (\ref{2-4}). Curves represent the functions $u(t_0,x)$ (blue),
$v(t_0,x)$ (red) and $w(t_0,x)$ (green)
   for  the fixed time $t_0=6$ and the parameters
  $\mu=\frac{3}{2}, \ d_1=d_3=2, \ \alpha_2=\alpha_4=5$
   (other parameters are calculated by formulae (\ref{2-5})). }\label{f2}
\end{figure}

\begin{figure}[ht]
\centering
 \includegraphics[width=6cm]{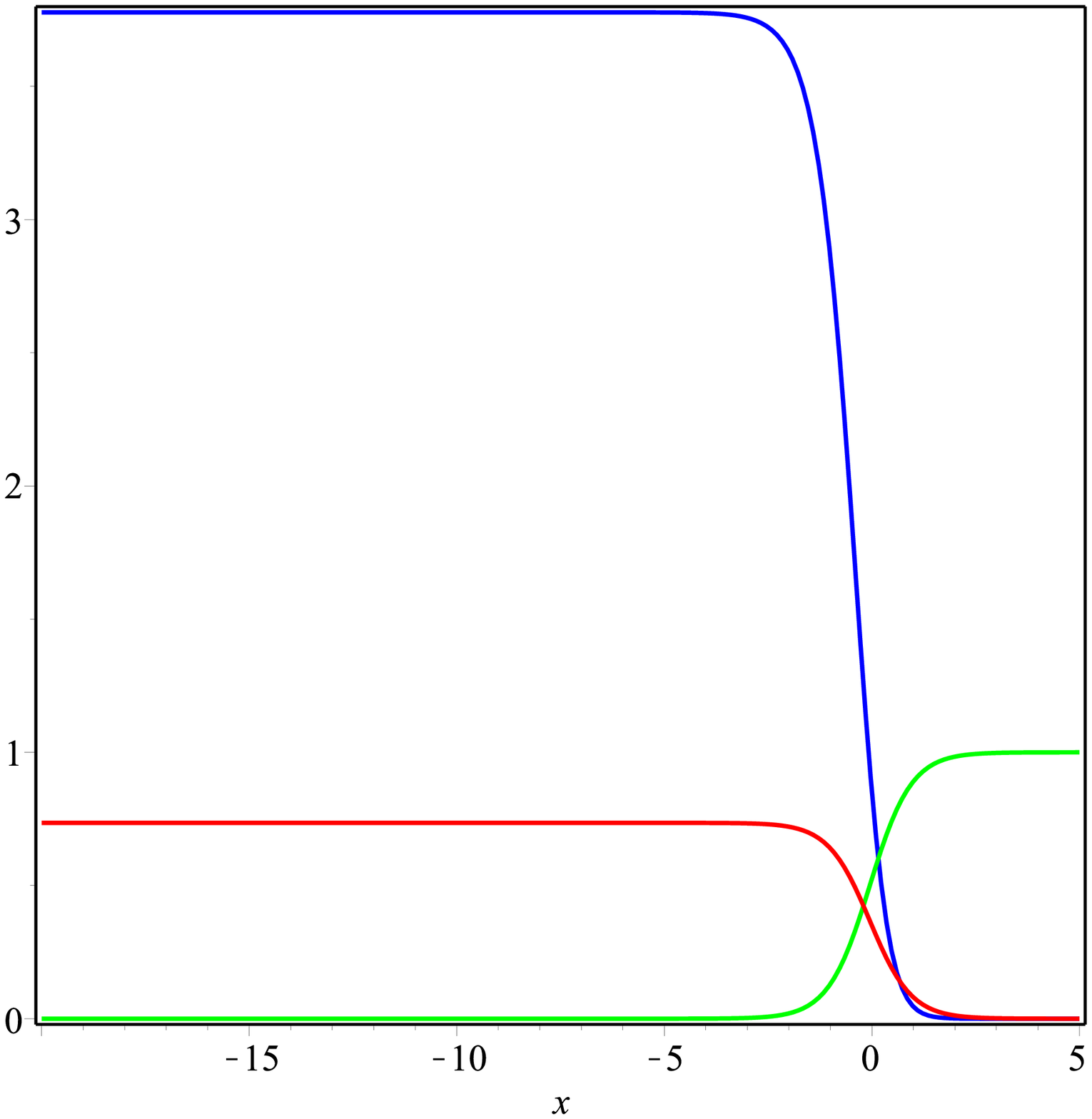}
  \includegraphics[width=6cm]{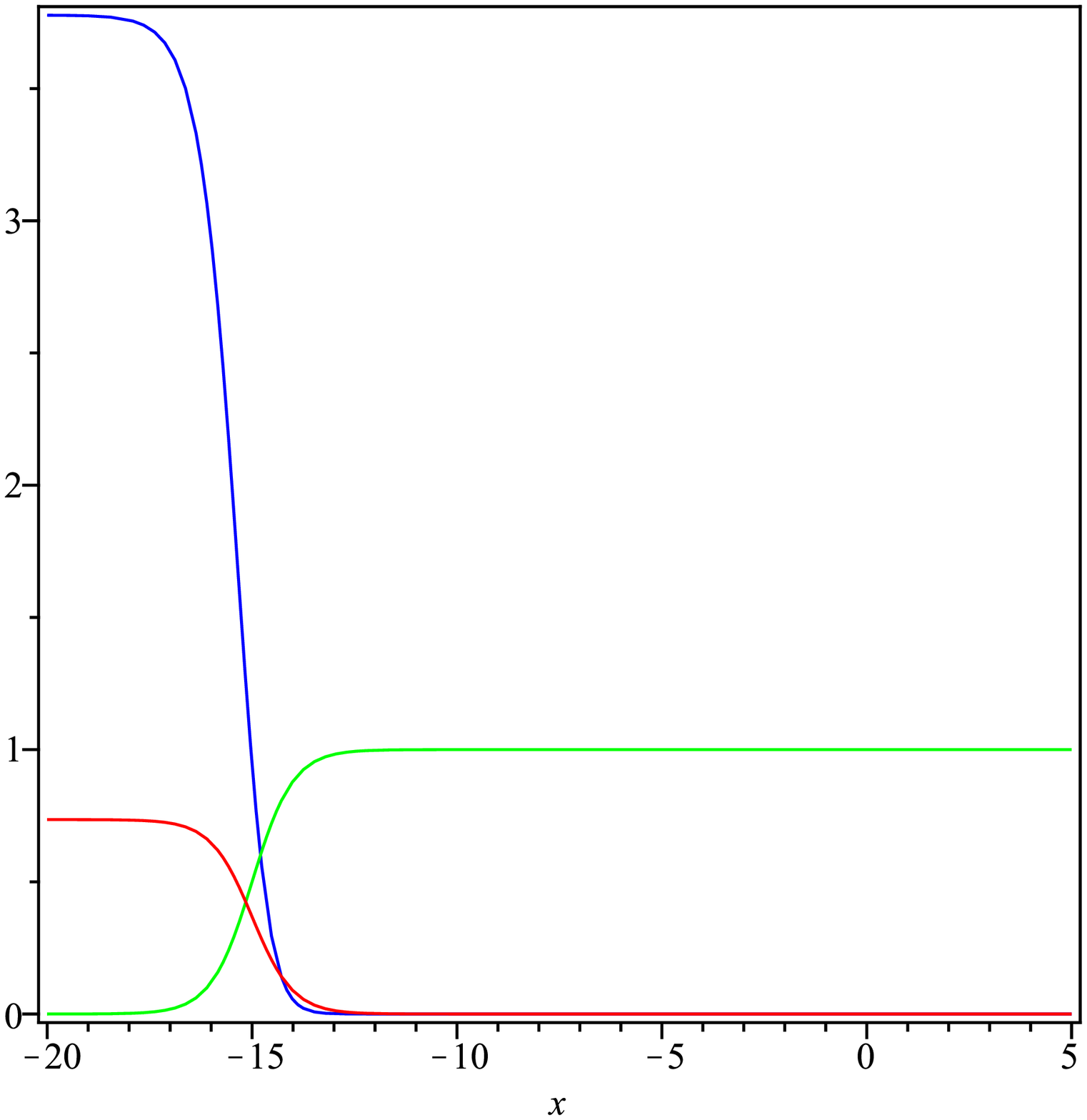}
 \caption{Traveling fronts (\ref{2-4}). Curves represent the functions $u(t_0,x)$ (blue),
$v(t_0,x)$ (red) and $w(t_0,x)$ (green)
   for the fixed time $t_0=0.01$ (left) and $t_0=3$ (right) and the parameters
  $\mu=-5, \ d_1=d_3=\frac{1}{2}, \ \alpha_2=\alpha_4=12$
   (other parameters are calculated by formulae (\ref{2-5})).
}\label{f3}
\end{figure}

Thus, we can use the formulae derived above in order to construct examples of traveling fronts, to plot the relevant curves (using the package Maple)  and to present their plausible interpretation. Figures \ref{f1}--\ref{f3} represent the exact solution (\ref{2-4}) in Case \textbf{(i)} $\mu>0$  (Fig.~\ref{f1}--\ref{f2})
and Case  \textbf{(ii)} $\mu<0$ (Fig.~\ref{f3}). All the curves satisfy the natural requirement of positivity at the given space intervals.

In Fig.~\ref{f1}--\ref{f2}, three traveling fronts are moving to the right along the $OX$ axes as it is predicted in Case \textbf{(i)}.  If we assume that the blue and green  curves represent  the communities  of  Russian language  speakers  and Ukrainian language  speakers, while the red curve describes the frequency of bilingual speakers, then the real language shift occurred in
  Ukraine during the Soviet period (from the end of the Second WW till the USSR  collapse) is  qualitatively described by these curves. In fact, the language situation in Ukraine can be approximated by the 1D model because the communities of different language  speakers varies very essentially from east to west (not so much from north to south).

 So, taking the point $(x=-4.0)$ as the eastern end and the point $(x=8.0)$ as the western end, one  realizes that the above curves at the time moment $t=0.01$ (see the curves in the left part of the figure) reflects the situation in  the end of the Second WW  (the  borders of the modern Ukraine were formed in that time). The frequency of   Russian language  speakers (blue curve)  was very high in the eastern part (see the interval $x \in [-4,-2]$), while an opposite situation was in the western part  (interval $x \in [6,8]$), in which  Ukrainian language dominated (actually the Russian language was unknown therein). In the central part of Ukraine (interval $x \in [-2,6]$), the linguistic  situation was more complicated  and this is shown in Fig.~\ref{f1} (left plot). However, one may say that  Ukrainian language  speakers (green curve) formed the main part of inhibitors  of the Central Ukraine and the frequency of using  this language decreased in the eastern direction. Finally, the community of bilingual speakers (red curve) was concentrated mostly in the east part after  the end of the Second WW.

 The time moment $t=4.0$ (see the curves in the right  part of the figure Fig.~\ref{f1}) reflects the situation in  the end of Soviet times, i.e. in the beginning  of 1990s. In that time, the community of    Russian language  speakers (blue curve) dominated in the east and  central part of Ukraine (interval $x \in [-4,6]$), the  community of bilingual speakers (red curve) was also strong in these parts.  However, the frequency of using  Ukrainian  language was very low and one may say about a rapid   extinction of this community. In that time, Ukrainian language dominated only in the western part of Ukraine, while there was also a part of  the Central Ukraine, in which the frequencies of using both languages was in some equilibrium (interval $x \in [4,6]$).

 Traveling fronts presented in  Fig.~\ref{f2} model the situation under the assumption that the USSR could exist 20--30 years  longer doing the same language politics, which was in favor of Russian language. Of course, one can expect the almost complete extinction of Ukrainian  language  speakers as it is shown (see green curve), however existence of a large community of bilingual speakers (red curve) seams to be not plausible. In fact, there is no any reason to study a `dead' language. So, we believe that the red curve  does not describe adequately  the frequency of using both languages for large values of time.

In Fig.~\ref{f3}, the exact solution (\ref{2-4}) is pictured in Case  \textbf{(ii)} $\mu<0$, so that the traveling fronts are moving to the left. As a result, the relevant interpretation is different. In fact, the time evolution leads to extinction of two communities, while only one monolingual  community is the winner of this language  competition.

 Finally, it should be pointed out that the exact solutions of the form  (\ref{2-4})  used above  for interpretation of the language shift occurred in
  Ukraine during the Soviet times present   do not   express  exact    numbers of speakers however these solution  describe  qualitatively the real linguistic situation.  In order to get adequate  quantitative  results, one needs to calculate correct coefficients in the RD system (\ref{1-2}) using census data in the former USSR. This is  another nontrivial problem, which will be treated elsewhere.

\section{Conclusions} \label{sec:4}

In this work, the  known three-component   reaction-diffusion system modeling the competition and co-existence  of two different language speakers  \cite{kandler-10} was a starting point. Such type competition leading to the  language shift occurs in many countries (territories)
and Ukraine is a typical example.
A modification of this system is proposed (see system (\ref{1-4})), which  was examined by the Lie
symmetry method. It was established that the system in question is invariant
 only w.r.t. the Lie operators of the time and space translations provided its  coefficient satisfy natural restrictions.
 Furthermore exact solutions in the form of  traveling
fronts are constructed using the tanh function  technique. As a result, four exact solutions in explicit form were found  for the first time.
One of them (see formulae  (\ref{2-4})) was studied in details  in order to identify its  properties.
 Having this done,
 plots of the traveling fronts were  drown  and the
 relevant interpretation describing the language shift occurred in
  Ukraine during the Soviet times was suggested.

  We are going to continue this  work. In particular,   some extension  of the model  is needed in order to take into account possible changes in language politics introduced by the government.

  Finally, it should be noted that a three-component model
 for describing the spread of an initially
localized population of farmers into a
 region occupied by hunter-gatherers  was introduced in \cite{ao-sh-shige-96} (see also the recent paper \cite{ch-dav-2019}, in which traveling fronts are constructed). It can be shown that the farmer--hunter-gatherers model can be derived
 from the RD system (\ref{1-2})  as a particular case.

\end{document}